# Majorana mode in vortex core of $Bi_2Te_3$/$NbSe_2$ topological insulator-superconductor heterostructure

Jin-Peng Xu[1‡], Mei-Xiao Wang[1‡], Zhi Long Liu[1], Jian-Feng Ge[1], Xiaojun Yang[2], Canhua Liu[1*], Zhu An Xu[2], Dandan Guan[1], Chun Lei Gao[1], Dong Qian[1], Ying Liu[1,3], Qiang-Hua Wang[4], Fu-Chun Zhang[2,5], Qi-Kun Xue[6], and Jin-Feng Jia[1*]

[1]Key Laboratory of Artificial Structures and Quantum Control (Ministry of Education), Department of Physics, Shanghai Jiao Tong University, Shanghai 200240, China
[2]State Key Laboratory of Silicon Materials and Department of Physics, Zhejiang University, Hangzhou 310027, China
[3]Department of Physics, Pennsylvania State University, University Park, PA 16802, USA
[4]National Laboratory of Solid State Microstructures and School of Physics, Nanjing University, Nanjing 210093, China
[5]Department of Physics, University of Hong Kong, Hong Kong, China
[6]State Key Laboratory for Low-Dimensional Quantum Physics, Department of Physics, Tsinghua University, Beijing 100084, China

**Majorana fermions have been intensively studied in recent years for their importance to both fundamental science and potential applications in topological quantum computing[1,2]. Majorana fermions are predicted to exist in a vortex core of superconducting topological insulators[3]. However, they are extremely difficult to be distinguished experimentally from other quasiparticle states for the tiny energy difference between Majorana fermions and these states, which is beyond the energy resolution of most available techniques. Here, we overcome the problem by systematically investigating the spatial profile of the Majorana mode and the bound quasiparticle states within a vortex in $Bi_2Te_3$/$NbSe_2$. While the zero bias peak in local conductance splits right off the vortex center in conventional superconductors, it splits off at a finite distance ~20nm away from the vortex center in $Bi_2Te_3$/$NbSe_2$, primarily due to the Majorana fermion zero mode. While the Majorana mode is destroyed by reducing the distance between vortices, the zero bias peak splits as a**

**conventional superconductor again. This work provides strong evidences of Majorana fermions and also suggests a possible route to manipulating them.**

† J.P.X. and M.X.W. contributed equally to this work.
* Corresponding authors: canhualiu@sjtu.edu.cn, jfjia@sjtu.edu.cn.

Identical to their antiparticles, Majorana fermions (MF) were proposed in 1937 as an alternative to Dirac theory of ordinary fermions that carry opposite charge from their antiparticles[4]. Neutrinos are the first candidate for MF in particle physics, but their Majorana status remains to be confirmed[5]. There are also proposals that quasiparticles in certain quantum condensed matter systems may be MFs. Examples include 5/2 fractional quantum Hall state, cold atoms and chiral p-wave superconductors[1,2]. Experimental realization of MFs is of great significance in fundamental physics. MFs obey non-Abelian statistics, and thus can be used to develop topological quantum computation. The recent work by Fu and Kane predicted that MFs should be present as zero-energy bound state at vortex cores of an engineered heterostructure consisting of a normal s-wave superconductor (SC) and a topological insulator (TI)[3]. Cooper pairs are introduced via proximity effect to TI surface where spin and momentum are locked in the topological surface state (TSS) band[6,7]. This leads to an unusual p-wave-like paired state that is time-reversal invariant and robust against disorder[8]. Theoretical studies later showed that the MFs may also reside at two ends of a semiconductor nanowire (NW) with strong spin-orbit coupling when it is contacted to an s-wave SC in a proper external magnetic field[9]. Several transport measurements revealed a signature of MFs, i.e., a sharp zero-bias peak in differential conductance spectrum, in various NW/SC junctions[10-13]. In InSb/Nb junction, an unconventional fractional a.c. Josephson effect was observed and attributed to the existence of MFs[14]. However, alternative explanations of these transport results based on disorder and/or band bending in the NWs have been proposed[15-18]. As yet, no conclusive evidence has been established for the existence of MF.

In contrary, the disorder alone is unlikely to induce a zero-bias peak in a TI/SC heterostructure, which can be used to detect MFs without complications mentioned above. Proximity effect induced superconductivity in a TI surface has been demonstrated in several TI/SC heterostructures[19-22]. To obtain direct evidence for the existence of MFs, a promising route is to detect the zero-bias bound state at vortex cores of a TI/SC heterostructure with scanning tunneling microscope and spectroscope (STM/STS), so that a single Majorana mode at a vortex core can be explicitly identified.

Very recently, we have succeeded in constructing TI/SC heterostructures with an atomically smooth interface by growing epitaxial thin films of $Bi_2Se_3$ and $Bi_2Te_3$ on $NbSe_2$ single crystals, where coexistence of Cooper pairs and TSS was illustrated[23,24]. Abrikosov vortices and Andreev bound states therein were observed in the $Bi_2Te_3/NbSe_2$ heterostructure with STM and STS[24]. The major difficulty to distinguish the zero mode MF in vortex core is the tiny energy gap separating the conventional quasiparticle states, which is estimated to be $0.83\Delta^2 / \sqrt{\Delta^2 + E_D^2}$, where $\Delta$ is the superconducting gap, $E_D$ is the Fermi energy relative to the Dirac point of the TSS band[25]. For $\Delta$ ~1 meV, $E_D$ ~100 meV, the mini-gap is ~0.01meV, which is much smaller than the present energy resolution (0.1meV) in STS. One way to increase the mini gap is to tune the Fermi level toward the Dirac point. However, in that case, the superconducting gap $\Delta$ becomes very small and the transition temperature becomes very low due to weaker proximity effect, hence the observation of Majorana mode is also difficult. In this work, we show that the spatial distributions of the Majorana mode and the bound states of a vortex can be used to identify the MF by STM.

Figure 1A is a schematic illustration for the configuration of the TI/SC heterostructure made by molecular beam epitaxy[26-27]. $Bi_2Te_3$ thin films grow on $NbSe_2$ in a layer-by-layer mode, resulting in very large atomically smooth terraces on the $Bi_2Te_3$ surface suited for vortices measurement by STS (the details on sample preparation is in the supplementary materials). As we shall describe below, the carriers of systems with 3 quintuple layers (QL) or less are almost from the bulk, and thus the vortex states are essentially the same as those in conventional s-wave superconductors. Systems with 5 or 6QL are topological insulators and the vortex states are expected to host MFs. We have simulated a single vortex for 5QL TI on top of a conventional s-wave SC. Our calculation shows that there is a pair of MFs, one at the surface and the other at the interface between the TI and the conventional SC, in the vortex core, as illustrated in Fig. 1B. We show the wavefunction amplitude (the electron part of the Nambu spinor) in a view field 100x100x5 of a lattice model of the 5QL TI. The amplitude is mainly concentrated on the top or bottom layers. The extent of the wavefunction is larger on the top layer, as we assumed that the proximity induced pairing potential is 50% smaller on the top layer than that on the bottom layer to cope with the experimental results (details for the numerics can be found in supplementary materials).

Figure 1C shows a typical contour of zero-bias differential conductance (ZBC) taken on a 1QL $Bi_2Te_3$ film in an external magnetic field of 0.1T. An Abrikosov vortex is clearly seen, which exhibits higher ZBC values due to the suppression of superconductivity within the vortex. At the center of the vortex, a peak in dI/dV due to the bound quasi-particle states can be measured as shown in Fig. 1D (see supplementary materials for the experimental conditions).

Along the dashed line indicated in Fig. 1C, we measured the spatial variation of the dI/dV spectra as a function of distance (r) away from the vortex center. The results are given in Fig. 2A. One can see that only one peak appears at zero-bias in the dI/dV spectra near the vortex center, and the peak splits into two at a finite distance r. The splitting energy increases linearly with r. For a better view, we plot dI/dV as functions of *r* and sample bias V in a fake color image in Fig. 2B, where the positions of the dI/dV peaks are indicated by red crosses. Two dotted lines are drawn to illustrate the linear relation between the energy of the split peaks and the distance r. Extrapolate the lines, the cross point also gives out the splitting start point. The results for 2 to 6QL $Bi_2Te_3$ films are shown in Figs. 2C-2G. Although the splitting can be resolved almost at the same position ~20nm from the center, the splitting start points (the cross points of the dotted lines) are obviously different for different films. For 1-3QL $Bi_2Te_3$ films, the peak splits right off the vortex center (zero-distance splitting), similar to that in a conventional s-wave superconductor, such as $NbSe_2$ (ref. 28). In contrast, for the thicker $Bi_2Te_3$ films (4-6QL), the splitting starts at a spatial point away from the vortex center (finite-distance splitting), an apparent deviation from that in a conventional superconductor. The peak splitting start position as a function of the thickness of $Bi_2Te_3$ films is summarized in Fig. 2H, a transition at 4 QL can be clearly observed.

The finite-distance splitting behavior of the bound states has not been reported before. We interpret this new feature related to the topological property of the local electronic structure. For the 4-6QL, the Fermi level lies near the top of the Dirac bands, and also crosses bottom of the bulk conduction bands (see Fig.4). The local density of states (LDOS) of a vortex as measured in our STM is contributed from both the bulk and the topological surface states. The bulk contribution is similar to that in a conventional superconductor, and the LDOS or the dI/dV spectra contributed from the bulk has a maximum (peak) at a final energy value proportional to the spatial distance r, see Fig. 2B for instance. In what follows we will argue that the MF mode of the 2D surface state may change the profile of the dI/dV spectra. For simplicity, we shall neglect the LDOS contribution from the quasi-particle bound states in 2D, since their contribution is expected to be similar to that from the bulk. The Majorana mode in the vortex core has been studied theoretically in Ref. 29. They calculated the LDOS for Nb/$Bi_2Se_3$/Nb sandwich structure, and showed that the MF mode has a spatial distribution of about 40 nm, with a sharp peak at zero bias in the dI/dV spectrum near the vortex core. Our sample structure has the similar parameters, so the spatial extension of the Majorana mode should be similar, although the envelope function depends on the Fermi wavevector. The Majorana mode is then expected to enhance the zero bias LDOS within a range of spatial distance r ~ 40 nm away from the vortex core, hence to possibly shift the maximum of the LDOS from a finite energy to zero bias energy for small r. The large zero bias LDOS at small r of the MF mode should be the underlying physics for the deviation of the zero distance splitting behavior of the bound state as we observed. Our STM measurement has an energy resolution of about 0.2meV. The LDOS within this energy resolution is expected to be enhanced due to the MF mode. We may argue that the maximum of LDOS for a fixed r may shift towards lower energy and the energy shift is less for larger r. These may explain the basic feature in our observation of the finite-distance splitting. Note that the effect of the MF mode to the change of the LDOS in the vortex core depends on the relative weight of the MF mode. A systematic study of the LDOS of a vortex with both the bulk and Dirac surface states will require further study. In brief, we interpret the observation of finite-distance split pattern as a clear demonstration of the MF in the center area of the vortex.

Our explanation is also supported by the magnetic field dependence of the LDOS of 5QL system. The dI/dV spectra taken at a vortex center of a 5QL $Bi_2Te_3$ film in various magnetic fields are shown in Fig. 3A. The zero-bias peak is very strong at a field less than 0.1T. As the field reaches 0.18T, the zero-bias peak becomes much weaker. In a conventional s-wave SC the vortex density is proportional to the magnetic field below a critical field $H_{c2}$. A single vortex structure is not sensitive to the external field and the LDOS near a vortex core is essentially unchanged as the field increases. The bound states of a vortex in $NbSe_2$ do not show the abrupt change when the magnetic field increases from 0.025T to 1.25T (Fig. 3B). The dramatic change in the zero-bias peak intensity in 5QL $Bi_2Te_3$ film is interpreted as the result of the coupling between adjacent vortices. At a small field, the distance between vortices is much larger than the vortex size, so the interaction between the vortices can be neglected. As the field increases to 0.18T, the distance between two adjacent vortices is reduced to about 110nm (Fig. 3C). Moreover, the higher magnetic field suppresses the superconducting gap and coherence length becomes longer, so the interaction between the vortices becomes strong enough to destroy the Majorana modes. In this case, the LDOS at vortex cores is governed again by conventional quasi-particle bound states and a zero-distance splitting pattern recovers in the spatial variation of dI/dV spetra. This situation is shown in Fig. 3D, which is in contrast to the finite-distance splitting at 0.1 T in Fig. 2F.

Although the finite-distance split pattern of the bound states peak may be explained by other reasons, e.g., the anisotropy of the superconducting gap[30], the split pattern changes with magnetic field can only be interpreted by existence of Majorana mode as far as we know. Therefore, the bound states splitting start point in this system is the indictor of the Majorana mode. Once the Majorana mode exists, the splitting start point is non-zero.

We now discuss the difference in electronic structure of various quintuple layer systems. Fig. 4 shows the evolution of the DOS near the Fermi level on $Bi_2Te_3$ films of various thicknesses obtained by STS. We can clearly see that TSS band or Dirac cone structure forms only when the thickness reaches 3QL. Once the TSS band forms, the DOS curve exhibits a deformed U-shape segment in the energy range between the bulk valence band maximum (VBM) and conduction band minimum (CBM)[31, 32]. In Fig. 4, arrows indicate the CBM, which shifts upward in energy as the thickness increases. It is seen that although the Dirac cone structure is formed at 3QL, the Fermi level is about 100meV higher than CBM, which means the bulk carrier density is still much higher than surface carrier density. So, the unconventional behavior from the TSS is submerged by that from bulk carriers. When the thickness reaches 4QL, the Fermi level is almost at the CBM, and for 5 and 6QL thin films, the Fermi level is a little bit lower than CBM. Therefore, the behavior related to TSS appears from 4QL, and the Majorana mode can only be clearly observed on 5 or 6 QL films grown on NbSe$_2$.

Strong evidences undoubtedly show that the Majorana mode existing in the vortex of the Bi$_2$Te$_3$/NbSe$_2$ hetero-structure. Since the TSS is protected by the time reversal symmetry, the Majorana mode in our configuration is free from impurities and defects. In fact, the defects may suppress the normal bound states in the vortex, and enhance the Majorana mode. It is also found the Majorana mode can be tuned off by increasing magnetic field. This might provide a route to controlling MF for quantum computation. In the future, it is highly interesting but challenging to pursue the quantum limit with the Fermi level near the Dirac point and $T < \Delta E$ (where $T$ is the temperature and $\Delta E$ is the energy level spacing of the bound states), in which the zero mode Majorana fermion may be directly observed.

**Acknowledgements**

JFJ thanks Shoucheng Zhang for reading the manuscript and helpful suggestions. The work at SJTU was supported by the National Basic Research Program of China (Grant No. 2012CB927401, 2011CB921902, 2013CB921902, 2011CB922200), NSFC (Grant No. 91021002, 11174199, 11134008, 11274228), Shanghai Committee of Science and Technology, China (No. 11JC1405000, 11PJ1405200, 12JC1405300). QHW was supported by NSFC (Grant No.11023002) and the Ministry of Science and Technology of China (Grant No.2011CBA00108 and 2011CB922101). FCZ was partially supported by NSFC (Grant No. 11274269) and RGC GRF (Grant No.707211)

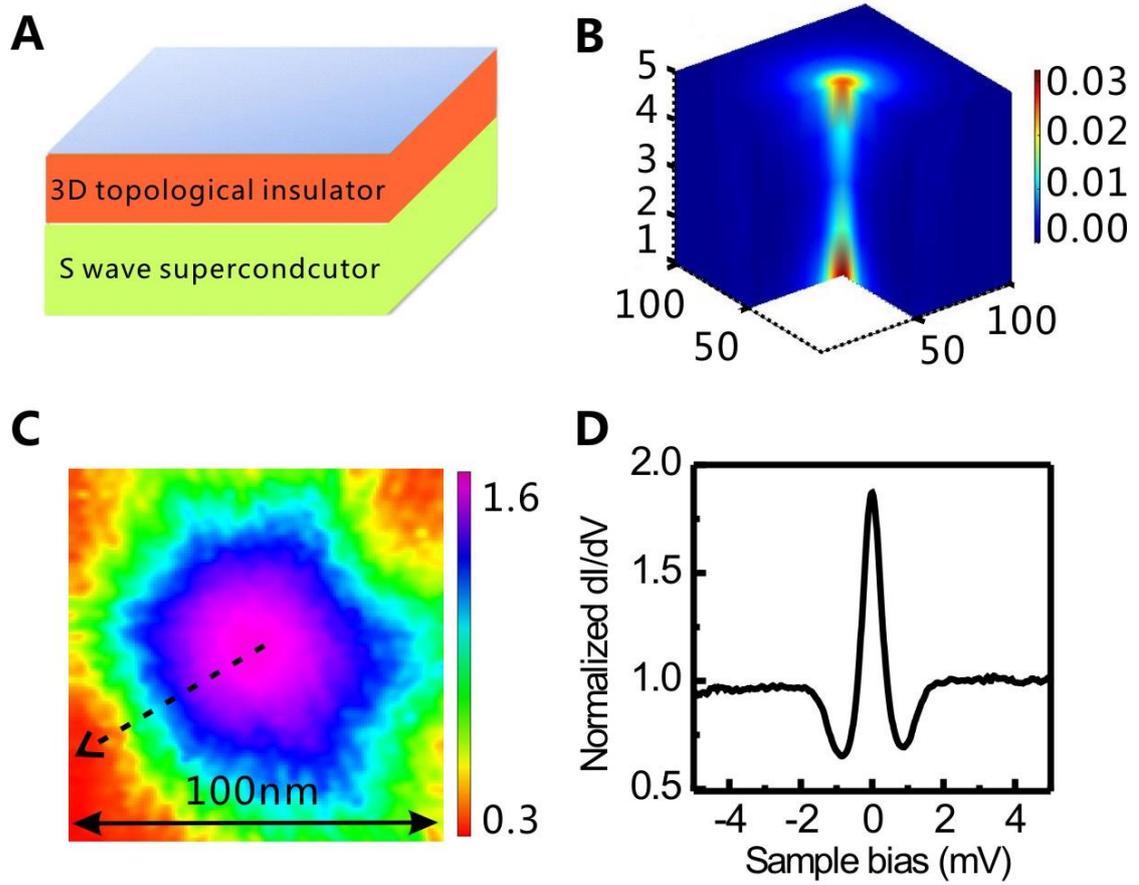

Fig. 1 (A) A schematic illustration of topological insulator/superconductor heterostructure. (B) The calculated results showing two Majorana modes in a vortex core on 5QL $Bi_2Te_3$/$NbSe_2$. (C) A vortex mapped by zero-bias dI/dV on 1QL $Bi_2Te_3$/$NbSe_2$ at 0.1T and 0.4K. (D) A sharp zero-bias peak in dI/dV spectrum measured at the center of the vortex in (C).

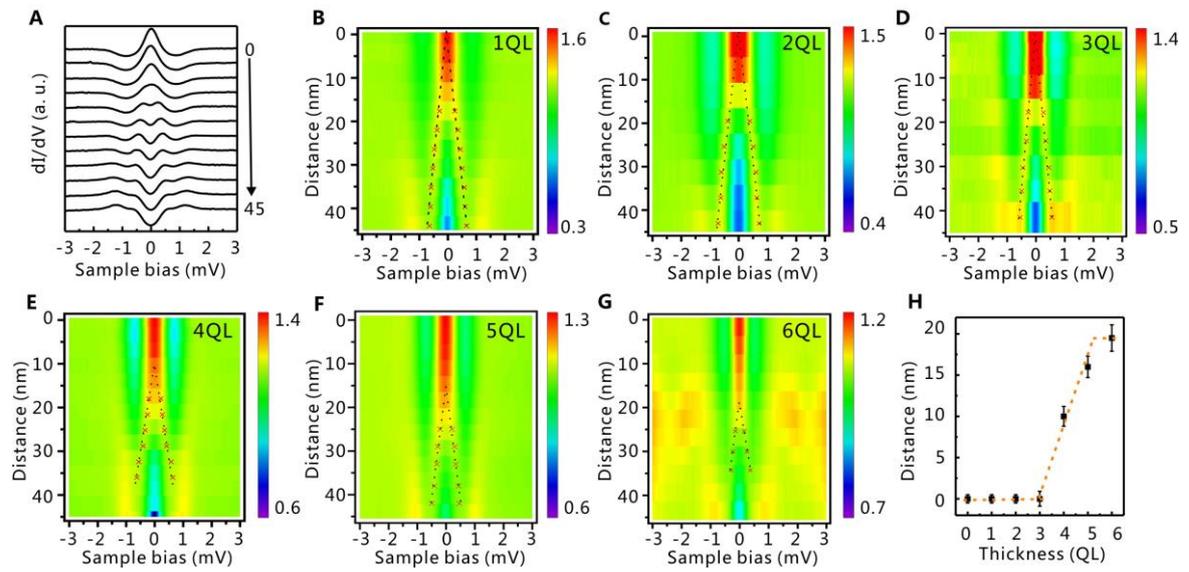

Fig. 2 (A) A series of dI/dV curves measured along the black dashed line in fig.1C, showing the peak of bound states splits into two at positions away from the vortex center. (B) The color image of (A) for a better view. The split peak positions in the dI/dV spectra are marked by red crosses, and the dotted lines superimposed on the crosses indicate start point of the peak splitting. (C-G) The experimental results for 2QL-6QL samples, following the similar data process of (B). (H) A summary of the start points of the peak split, showing a crossover at 4QL.

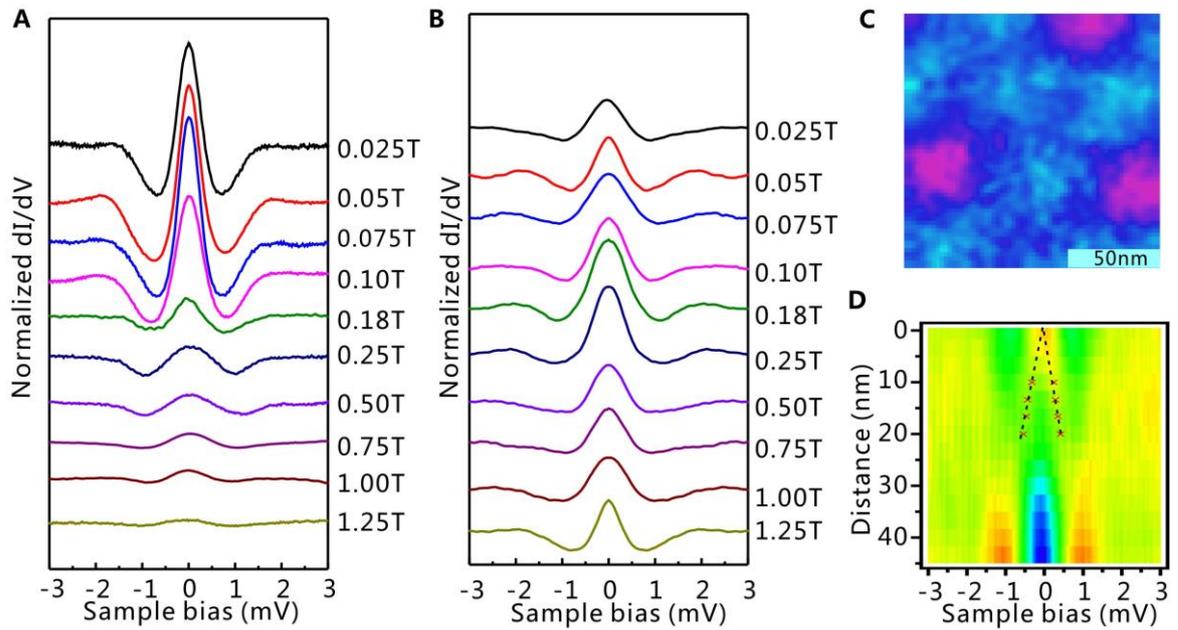

Fig. 3 A series of dI/dV spectra measured at various magnetic fields at a vortex center of 5QL $Bi_2Te_3$/$NbSe_2$ (A) and bare $NbSe_2$ (B). A dramatic drop in the peak intensity is clearly seen at 0.18T in (A). (C) Vortex lattice measured at 0.18T on the 5QL $Bi_2Te_3$/$NbSe_2$. The average distance between two adjacent vortices is about 110nm. (D) The color image of dI/dV spectra measured at the vortex cores shown in (C). The crosses and dotted lines are the same indicators as those in Fig. 2(B). The peak-splitting start point is zero, in sharp contrast to that in fig.2(F).

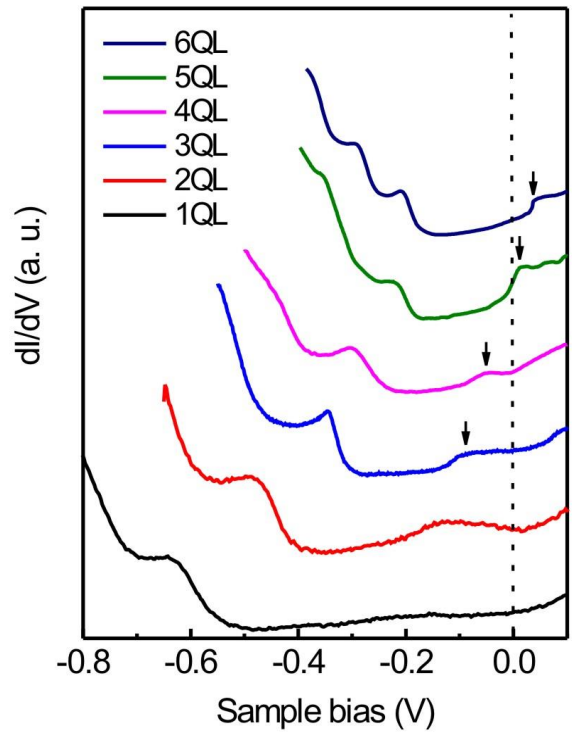

Fig. 4 A series of dI/dV spectra measured on 1QL-6QL Bi$_2$Te$_3$ films on NbSe$_2$ at 0.4 K. The curves become a deformed U-shape after the Dirac cone structure forms at 3QL. The black arrows indicate the energy position of CBM, and the dashed line indicates the Fermi energy.